\documentclass[slac_one]{revtex4}
\usepackage{graphicx}
\usepackage{fancyhdr}
\begin{document}
\title{$B \to \pi$ and $B_{(s)} \to K$ form factors and $V_{ub}$ determination} 

%

\author{Bla\v zenka Meli\'c}
\affiliation{Rudjer Bo\v skovi\'c Institute, Theoretical Physics Division, P.O.Box 180, HR-10000 Zagreb. Croatia}

\begin{abstract}
We present a QCD light-cone sum rule (LCSR) estimation of the $B \to \pi$, $B \to K$, and 
$B_s \to K$ form factors calculating gluon radiative corrections at next-to-leading order. 
The $\overline{MS}$ b-quark
mass is used, instead of the one-loop pole mass employed in the previous
analyses. For $B \to K$ and $B_s \to K$ form factors the SU(3)-symmetry breaking
corrections are included, both in the hard-scattering kernels and in the distribution amplitudes (DAs). 
By combining the predicted value for $f^+_{B\pi}(0)$
with the product $|V_{ub} f^+_{B\pi}|$ extracted from the $B \to \pi l\nu_l$ measurement, we obtain
the LCSR prediction for $V_{ub}$ CKM matrix element.
\end{abstract}

\maketitle

\thispagestyle{fancy}


\section{INTRODUCTION} 

Nowadays the measurements in the flavor physics are mostly dedicated to overdetermination of 
the unitarity triangle of the CKM matrix. 
One of the sides of the triangle is given by the CKM matrix element $V_{ub}$. It can be determined from inclusive or 
exclusive semileptonic $B$ decays, which are complementary in a sense that they involve different theoretical 
(and experimental) methods for $V_{ub}$ extraction. 
The inclusive $V_{ub}$ determination heavily relies on an accurate calculation of the decay spectrum under 
stringent kinematical cuts.  On the other hand, the 
$V_{ub}$ extraction from the exclusive semileptonic $B \to \pi l \nu_l$ decay requires the knowledge of the $B \to \pi$ form factor, 
$f_{B\pi}^+$, which is determined by nonperturbative methods, either by lattice calculations, or by applying QCD sum rules. 

Moreover, the $B \to \pi$ and $B_{(s)} \to K$ form factors serve as the main ingredients of different factorization 
models for calculating hadronic matrix elements in two-body nonleptonic $B$ decays, as one can see below:
\begin{eqnarray}
\langle \pi\pi | {\cal O}_1 | B \rangle 
= \underbrace{\langle \pi |\overline{d}\Gamma_{\mu}u| 0 \rangle
\langle \pi |\overline{u}\Gamma^{\mu}b| B \rangle }_{'naive' \; factorization}
\Big [ 1 +  O(\alpha_s, \Lambda_{QCD}/m_b) \Big ]  
= i m_b^2 f_{\pi} f_{B\pi}^+(m_{\pi}^2)
\Big [ 1 + O(\alpha_s, \Lambda_{QCD}/m_b) \Big ] \, , 
\end{eqnarray}
since they enter already at the leading level of a calculation. 

The estimation of the $SU(3)$ violation among the  form factors is important for assessing the validity of various isospin and SU(3) 
relations applied to constrain new physics contributions. For example, in the relation 
\begin{eqnarray}
A(B^- \rightarrow \pi^-\overline{K}^0) +
\sqrt{2} A(B^- \to \pi^0 K^-) 
= \sqrt{2} \frac{V_{us}}{V_{ud}} A(B^- \to \pi^- \pi^0)
( 1 + \Delta_{SU(3)})
\end{eqnarray}
$\Delta_{SU(3)}$ measures the net SU(3) breaking effect which comes from ratio of the form factors $f_{BK}/f_{B\pi}$, 
the decay constants $f_K/f_{\pi}$, etc. 

Therefore, here we intend to explore the $B_{(s)} \to \{\pi,K\}$ form factors in details by using the light-cone sum rules (LCSRs).

\section{FORM FACTORS FROM LIGHT-CONE SUM RULES}

Heavy-to-light vector, $f_{B_{(s)}P}^+$, and scalar $f_{B_{(s)}P}^0 = f^+_{B_{(s)}P}(q^2) + q^2/(m_B^2 - m_{P}^2) 
f^-_{B_{(s)}P}(q^2)$, form factors originate from the relation
\begin{eqnarray}
\langle P(p)|\bar{q} \gamma_\mu b |\bar B_{(s)}(p+q)\rangle &=&
2 f^+_{B_{(s)}P}(q^2)p_\mu +\left[ f^+_{B_{(s)}P}(q^2)+f^-_{B_{(s)}P}(q^2) \right] q_\mu\,,
\end{eqnarray}
while the penguin form factor, $f_{B_{(s)}P}^T$ is defined as 
\begin{eqnarray}
\langle P(p)|\bar{q} \sigma_{\mu \nu}q^\nu b |\bar B_{(s)}(p+q)\rangle &=&
\Big [q^2(2p_\mu+q_\mu) - (m_{B_{(s)}}^2-m_P^2) q_\mu\Big ]
\frac{i f_{B_{(s)}P}^T(q^2)}{m_{B_{(s)}}+m_P}\,.
\end{eqnarray}
In the above relations $P = \pi$ or $K$, and $q=u$ for  $B \to \pi$, $q=s$ for $B \to K$ and $q = d$ for $B_s \to K$ transitions. 
To obtain these form factors from the LCSR one introduces the correlator with the vector and the penguin current, 
\begin{eqnarray}
& & i\int d^4x ~e^{i q\cdot x}
\langle P(p)|T\left\{\bar{q}(x)\Gamma_\mu b(x), j_{B_{(s)}}(0)\right\}|0\rangle
= \Bigg\{\begin{array}{ll}
F_{(s)}(q^2,(p+q)^2)p_\mu +\widetilde{F}_{(s)}(q^2,(p+q)^2)q_\mu\,,& ~~\Gamma_\mu= \gamma_\mu\\
&\\
F^T_{(s)}(q^2,(p+q)^2)\big[p_\mu q^2-q_\mu (q p)\big]\,,& ~~\Gamma_\mu= -i\sigma_{\mu\nu}q^\nu\\
\end{array}
\label{eq:corr}
\end{eqnarray}
where the interpolating currents for the $B$ and $B_s$ mesons are $j_B = m_b\bar{b}i\gamma_5 d$ and 
$j_{B_s} = (m_b+m_s)\,\bar{b}i\gamma_5 s$, 
respectively. The light quark masses $m_{u,d}$ are neglected. 
For large virtualities of the currents in (\ref{eq:corr}), the correlator is dominated by the light-cone distances, $x^2 \to 0$. 
Therefore one is allowed to perform the light-cone OPE, in terms of the light-cone DAs of increasing twist. 
By using dispersion relations and the quark-hadron 
duality assumption, the correlator is related to the sum over hadronic states which is proportional to 
$f_{B_{(s)}}f_{B_{(s)}P}^{+,0,T}$. For $B \to \pi$ vector form factor the final expression has the form
\begin{eqnarray}
\frac{2m_B^2 f_Bf^+_{B\pi}(q^2)}{m_B^2-(p+q)^2}
= \frac{1}{\pi} \int_{m_b}^{s_0^B} \frac{ds}{s - (p + q)^2}
\sum_{n=twist}  \int_0^1 du {\rm Im}_s T_H^{(n)} \Phi_{\pi}^{(n)}\,.
\end{eqnarray}
In addition, one needs the Borel transformation 
$ 1/(s - (p + q)^2)^n \stackrel{s = (p+q)^2} \Rightarrow  1/(M^2)^n e^{-s/M^2}/\Gamma(n)$ 
in order to suppress higher states and to enhance the ground state contribution. 
The Borel parameter $M^2$ and the effective continuum threshold parameter $s_0^B$ are parameters which have to be fixed by 
following certain criteria as explained in details in \cite{DKMMO,DM}. Since the decay constant $f_B$ can be also estimated by 
the sum rules, one can consistently perform estimation of the form factors and reduce the parameter uncertainties. 
In (6), $T_H^{(n)}$ is perturbatively calculable hard-scattering part and $\Phi_{\pi}^{(n)}$ is the light-cone distribution amplitude 
of twist $n$. The leading twist-2, two-particle DA $\phi_{\pi}$ is defined as
\begin{eqnarray}
\langle \pi (q) | \overline{u}(x) \gamma_{\mu} \gamma_5 d(0) | 0 \rangle_{x^2 =0}
= -i q_{\mu} \frac{f_{\pi}}{\sqrt{2}} \int_0^1 du e^{ i u q \cdot x} \phi_{\pi}(u,\mu)\, ,
\end{eqnarray}
where $u$ is the fraction of the momentum carried by a meson's constituent. In general there is a 
Gegenbauer polynomial expansion of the leading twist-2 DAs
\begin{eqnarray}
\phi_{\pi,K}(u,\mu) = 6 u (1-u) \left \{
1 + \sum_{n=0}^{\infty} a_n^{\pi,K}(\mu) C_n^{3/2}(2 u -1) \right \} \stackrel{\mu \to \infty}{=} 6 u (1-u)
\end{eqnarray}
with the well-known asymptotic behavior. The coefficients in the expansion are determined either by combining the sum 
rules for the pion form factors with experimental data or from 
lattice calculations. For the pion DA:  $a_1^{\pi}(\mu) = 0$, $a_2^{\pi}(\mu)= 0.25 \pm 0.15$, 
$a_4^{\pi}(\mu) \simeq 0$, and for the kaon: $a_1^K(\mu) = 0.10 \pm 0.04$, $a_{2,4}^{K} \simeq a_{2,4}^{\pi}$. 
The structure of pseudoscalar and vector DAs is know to twist-4 accuracy. 
At the higher twists there exists  two- and three-particle DAs. They are related by Wandzura-Wilczek-type relations and their 
parameters are obtained from the two-point sum rules or by some models.

\section{$B \to \pi$ FORM FACTORS AND DETERMINATION OF $V_{ub}$ \cite{DKMMO}}

In the calculation \cite{DKMMO} we employ the $\overline{MS}$ scheme, and use 
$\overline{m}_b( \overline{m}_b)= 4.164 \pm 0.025 \; \mbox{GeV}$. 
The decay constants are calculated in the same scheme by using the two-point sum rule from \cite{JL} and 
at $O(\alpha_s)$ for our values of parameters we obtain
\begin{equation}
f_B = 214 \pm 18 \; {\rm MeV} , \,  f_{B_s} =  250 \pm 20 \; {\rm MeV}.
\end{equation}
The sum rule parameters: the scale $\mu$, the Borel parameter $M$ and the threshold parameter $s_0^B$ are estimated by 
taking all other parameters, specified in \cite{DKMMO,DM}, at their central values and allowing the coefficients of the 
leading twist-2 DA to vary within their intervals. In addition, 
we require that the subleading twist-4 terms in the LO are small,
less than $3\%$ of the LO twist-2 term, that the NLO corrections of twist-2 and twist-3 parts are not exceeding $30\%$ of 
their LO counterparts, and that the subtracted continuum remains
small, which fixes the allowed range of $M^2$. The effective threshold parameters are fitted so that 
the derivative over $-1/M^2$ of the expression of the complete LCSRs for a particular form factor 
reproduces the physical mass $m_{B}^2$ with a high accuracy of $O(0.5 \%-1 \%)$ in the stability region of the sum rules. 
These demands provide us the following central values for the sum rule parameters: $\mu = 3 \, {\rm GeV}$, $M^2 = 18\,{\rm GeV}^2$, 
$s_0^B = 35.75 \,{\rm GeV}^2$.  The predicted vector $B \to \pi$ form factor at zero momentum transfer then reads
\begin{equation}
f^+_{B\pi}(0)= 0.263 ^{~+ 0.004}_{~-0.005}\bigg|_{M,\overline{M}}
\,_{-0.004}^{+0.009}\bigg|_{\mu} \pm 0.02 \bigg|_{shape}
\,^{+0.03}_{-0.02}\bigg|_{\mu_\pi}\pm 0.001\bigg|_{m_b} \, , 
\end{equation}
while  its $q^2$ dependence is depicted on Fig.1. The first error comes from the uncertainties in the Borel parameters for 
$f_{B\pi}^+$ $(M)$ and $f_B$ $(\overline{M})$. The largest uncertainties are due to variation of the quark masses in $\mu_{\pi} 
= m_{\pi}^2/(m_u + m_d)$ and due to the fitting of the experimental shape by varying of $a_2^{\pi}$ and 
$a_4^{\pi}$ twist-2 DA parameters. 
\begin{figure}
\begin{center}
\includegraphics[width=8cm]{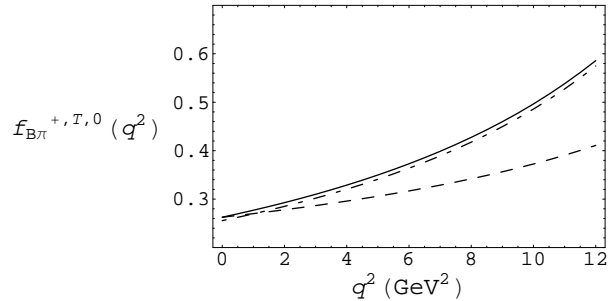}
\end{center}
\caption{ \it The LCSR prediction for 
form factors $f^+_{B\pi}(q^2)$ (solid line), $f^0_{B\pi}(q^2)$ (dashed line) 
and $f^T_{B\pi}(q^2)$ (dash-dotted line) 
at $0<q^2<12$ GeV$^2$ and for the central values of all input parameters. }
\end{figure}
Making the same numerical analysis for the penguin form factor and adding all uncertainties in the quadratures we finally predict
\begin{eqnarray}
f_{B\pi}^0(0) = f_{B\pi}^{+}(0)  = 0.26^{+0.04}_{-0.03} \, , \qquad 
f_{B\pi}^T(0) = 0.255 \pm 0.035 \, .
\end{eqnarray}
Obtained results are close to other LCSR and lattice results on $B \to \pi$ form factors. 

The $V_{ub}$ matrix element is extracted from the P.Ball's fit \cite{Ball} of $|V_{ub} f^+_{B\pi}|$  to BaBar data on
$B \to \pi l \nu_l$
and amounts to 
\begin{equation}
|V_{ub}|=\bigg(3.5 \pm 0.4\big|_{th} 
\pm 0.2\big|_{shape}\pm 0.1\big|_{BR}\bigg)\times 10^{-3}\,,
\label{eq:Vub}
\end{equation}
where the first error is due to the estimated uncertainty 
of $f^+_{B\pi}(0)$  and the two remaining 
errors  originate from the experimental errors of $|V_{ub} f^+_{B\pi}|$.
The prediction is in agreement with other recent determinations of $V_{ub}$ from exclusive $B \to \pi l \nu_l$ decay, see 
Table 1 in  \cite{DKMMO}.

\section{$B_{(s)} \to K$ FORM FACTORS AND SU(3) BREAKING EFFECTS \cite{DM}}

In the analysis of the $B_{(s)} \to K$ form factors \cite{DM} we include, apart from the 
SU(3) breaking effects of  the parameters of the leading twist DAs, such are $f_K/f_{\pi}$ and $\mu_K/\mu_{\pi}$,  
the complete SU(3)-symmetry breaking corrections in the $K$ meson DAs  \cite{BBL}  for all twist-3 and twist-4 two- and
 three-particle DAs. In the hard-scattering amplitudes at LO we consider $p^2 =m_K^2$ corrections. 
At next-to-leading order (NLO) in the hard-scattering amplitudes, the mass effects cause nontrivial mixing between 
twist-2 and twist-3 DAs. 
Therefore, at NLO in the hard-scattering parts we set $p^2=m_K^2 =0$, and consistently use twist-2 and twist-3 two-particle 
kaon DAs without mass corrections. However, we analyze and include the kaon mass effects  in the error estimates. 

Since the LO hard-scattering amplitudes are already complicated when the twist-4 and three-particle DAs are included, and the mass 
effects make the calculation even more demanding, the work can be greatly simplified by applying the method of numerical 
integration of the amplitudes in the complex plane, making the usual analytical extraction of the imaginary parts of 
LCSR hard-scattering amplitudes obsolete.
%
%
With the same conditions for the fitting of the sum rule parameters as those applied  in the $B \to \pi$ calculation, 
we extract the central values, 
$\mu = 3\, {\rm GeV}$, $M^2 = 18.0 \, {\rm GeV}$ and $s_0^B = 38 \, {\rm GeV}$ and obtain
\begin{eqnarray}
f_{BK}^+(0) = f_{BK}^{0}(0)  = 0.36^{+0.05}_{-0.04}\,,  \qquad
f_{BK}^T(0) = 0.38\pm 0.05 \, .
\end{eqnarray}
The $B_s\to K$ form factors are estimated at the default values $\mu = 3.4\, {\rm GeV}$, $M^2 = 19.0 \, {\rm GeV}$ and 
$s_0^B = 39 \, {\rm GeV}$ and the results are 
\begin{eqnarray}
f_{B_sK}^+(0) = f_{B_sK}^{0}(0)  = 0.30^{+0.04}_{-0.03}\,,  \qquad
f_{B_sK}^T(0) = 0.30\pm 0.05 \, .
\end{eqnarray}

Having the predictions for the $B \to \pi$  and $B_{(s)}K$ form factors calculated in the same model, 
we are able to get the SU(3)-breaking corrections which amount to relatively large SU(3)-breaking corrections in 
$B \to K$ decays, and somewhat smaller in $B_s \to K$ decays:
\begin{eqnarray}
\frac{f_{BK}^+(0)}{f_{B\pi}^+(0)} &=& 1.38^{+0.11}_{-0.10}\,,  \qquad\qquad\qquad \frac{f_{B_sK}^+(0)}{f_{B\pi}^+(0)} = 1.15^{+0.17}_{-0.09}\,,
\\
\frac{f_{BK}^T(0)}{f_{B\pi}^T(0)} &=& 1.49^{+0.18}_{-0.06}\,,  \qquad\qquad\qquad \frac{f_{B_sK}^T(0)}{f_{B\pi}^T(0)} = 1.17^{+0.15}_{-0.11}\,.
\end{eqnarray}

By checking some of the SU(3) and U-spin relations in the factorization models for $B_{(s)} \to K \pi, K K $ amplitudes like 
\begin{equation}
\xi = \frac{f_K}{f_{\pi}}\frac{f^+_{B\pi}(m_K^2)}{f_{B_sK}^+ (m_{\pi}^2)}\frac{m_B^2 - m_{\pi}^2}{m_{B_s}^2 - m_K^2} =
1.01^{+0.07}_{-0.15} \, , 
\end{equation}
and 
\begin{equation}
\frac{A_{fact}(B_s \to K^+K^-)}{A_{fact}(B_d \to \pi^+ \pi^-) }
= \frac{f_K}{f_{\pi}} \frac{f_{B_sK}^+ (m_{K}^2)}{f^+_{B\pi}(m_{\pi}^2)} \frac{m_{B_s}^2 - m_K^2}{m_B^2 - m_{\pi}^2} =
1.41^{+0.20}_{-0.11} \, , 
\end{equation}
we have found that SU(3) and U-spin relations are case by case badly broken and therefore, for each particular case have to be
 carefully examined. 

\begin{acknowledgments}
The author acknowledges the collaboration with G. Duplan\v ci\'c, A. Khodjamirian, Th. Mannel and N. Offen on the subject 
presented here.  The work is supported by the
Ministry of Science, Education and Sport of the
Republic of Croatia, under contract 098-0982930-2864 and partially by Alexander von Humboldt foundation.
\end{acknowledgments}

\end{document}